\documentclass[10pt,letterpaper,aps,prl,twocolumn,superscriptaddress]{revtex4-1}
\usepackage{amsmath}
\usepackage{amsfonts}
\usepackage{amssymb}
\usepackage{bm}
\usepackage{MnSymbol,wasysym}
\usepackage[pdftex]{graphicx}
\usepackage{epstopdf}
\usepackage{etoolbox}
\usepackage{times}
\usepackage[letterpaper,margin=1in]{geometry}
\usepackage[utf8]{inputenc}
\usepackage{color}
\usepackage[usenames,dvipsnames]{xcolor}
\usepackage{epstopdf}
\usepackage{units}
\usepackage[normalem]{ulem}
\usepackage{enumerate}
\usepackage[caption=false]{subfig}
\usepackage{longtable}
\usepackage{float}
\usepackage{braket}
\usepackage{hyperref}
\usepackage{silence}
\definecolor{Red}{rgb}{1,0,0}
\begin{document}
\title{Observation of a large, resonant, cross-Kerr nonlinearity in a free-space Rydberg medium}

\author{Josiah Sinclair}
\affiliation{Department of Physics, and Centre for Quantum Information and Quantum Control,  University of Toronto, 60 St. George Street, Toronto, Ontario, Canada M5S 1A7}
\author{Daniela Angulo}
\affiliation{Department of Physics, and Centre for Quantum Information and Quantum Control,  University of Toronto, 60 St. George Street, Toronto, Ontario, Canada M5S 1A7}
\author{Noah Lupu-Gladstein}
\affiliation{Department of Physics, and Centre for Quantum Information and Quantum Control,  University of Toronto, 60 St. George Street, Toronto, Ontario, Canada M5S 1A7}
\author{Kent Bonsma-Fisher}
\affiliation{Department of Physics, and Centre for Quantum Information and Quantum Control,  University of Toronto, 60 St. George Street, Toronto, Ontario, Canada M5S 1A7}
\affiliation{National Research Council of Canada, 100 Sussex Drive, Ottawa, Ontario, Canada K1A 0R6}
\author{Aephraim M.  Steinberg}
\affiliation{Department of Physics, and Centre for Quantum Information and Quantum Control,  University of Toronto, 60 St. George Street, Toronto, Ontario, Canada M5S 1A7}
\affiliation{Canadian Institute For Advanced Research, 180 Dundas St. W., Toronto, Ontario, Canada, M5G 1Z8}

\date{\today}
\begin{abstract}
We report the experimental observation of a cross-Kerr nonlinearity in a free-space medium based on resonantly-excited, interacting Rydberg atoms and electromagnetically induced transparency. The nonlinearity is used to implement cross-phase modulation between two optical pulses. The nonlinear phase written onto the probe pulse is measured to be as large as $8$ mrad per nW of signal power, corresponding to a $\chi^{(3)}$ of $10^{-8}\textrm{m}^2/\textrm{V}^2$. Potential applications range from optical quantum information processing to quantum non-demolition measurement of photon number.
\end{abstract}


\maketitle

Fantastically strong interactions between high-lying Rydberg states combined with electromagnetically induced transparency (EIT) are an extremely promising platform for nonlinear optics at the level of single photons. In the last decade, Rydberg atoms have been used to demonstrate phase gates \cite{Tiarks2018}, photonic switches \cite{Baur2014a}, atomic logic gates \cite{Muller2009,Keating}, quantum memory \cite{Distante2016,Maxwell2013}, the generation of single-photon Fock states \cite{Dudin2012a,Ripka446}, and other non-classical states of light \cite{Firstenberg2013,Peyronel2012c,Thompson2017,Liang2018}. What these experiments share in common is that the nonlinear effects observed saturate at or near one photon. This is in contrast to a Kerr nonlinearity, where a medium experiences a \textit{linear} shift in the index of refraction proportional to the number of photons in the medium. The Kerr effect has been studied extensively and has well-known applications in optical quantum computing \cite{NemotoMunro2004,MunroNemoto2005} and in generating and measuring non-classical states of light \cite{Imoto1985,Schmid2017}. In light of this, there have been several theoretical studies endeavoring to harness the powerful interactions of Rydberg atoms to implement a Kerr nonlinearity \cite{Bai2008,Pohl2011a,Pohl2011b,Grangier2013,Grangier2014,grankin2014,Grankin2015, Bienias2016}. Noteworthy experimental results include the observation and study of a large dissipative (imaginary) $\chi^{(3)}$ \cite{Pritchard2010,Pritchard2011a, Boddeda2016Experiment} and the first observation of an off-resonant, dispersive, self-Kerr nonlinearity, which was used to demonstrate cavity-enhanced self-phase modulation \cite{Parigi2012,Boddeda2016Experiment}. Notably, all experiments to date have involved a single beam, whereas most applications, such as photon number squeezing or quantum non-demolition measurement of photon number, require a cross-Kerr effect involving two beams.

In this letter, we report the first experimental observation of a dispersive cross-Kerr nonlinearity based on resonant Rydberg EIT. We observe that the phase shift acquired by a resonant optical pulse propagating through a cold cloud of atoms under EIT conditions depends linearly on the intensity of a second optical pulse. We present a simple theoretical treatment based on van der Waals interactions, which provides an intuitive explanation for the origin and scaling of the observed cross-phase shifts and is consistent with our observations.



\begin{figure}[h]
        \centering
        \resizebox{8cm}{!}{
                \includegraphics[width=\textwidth]{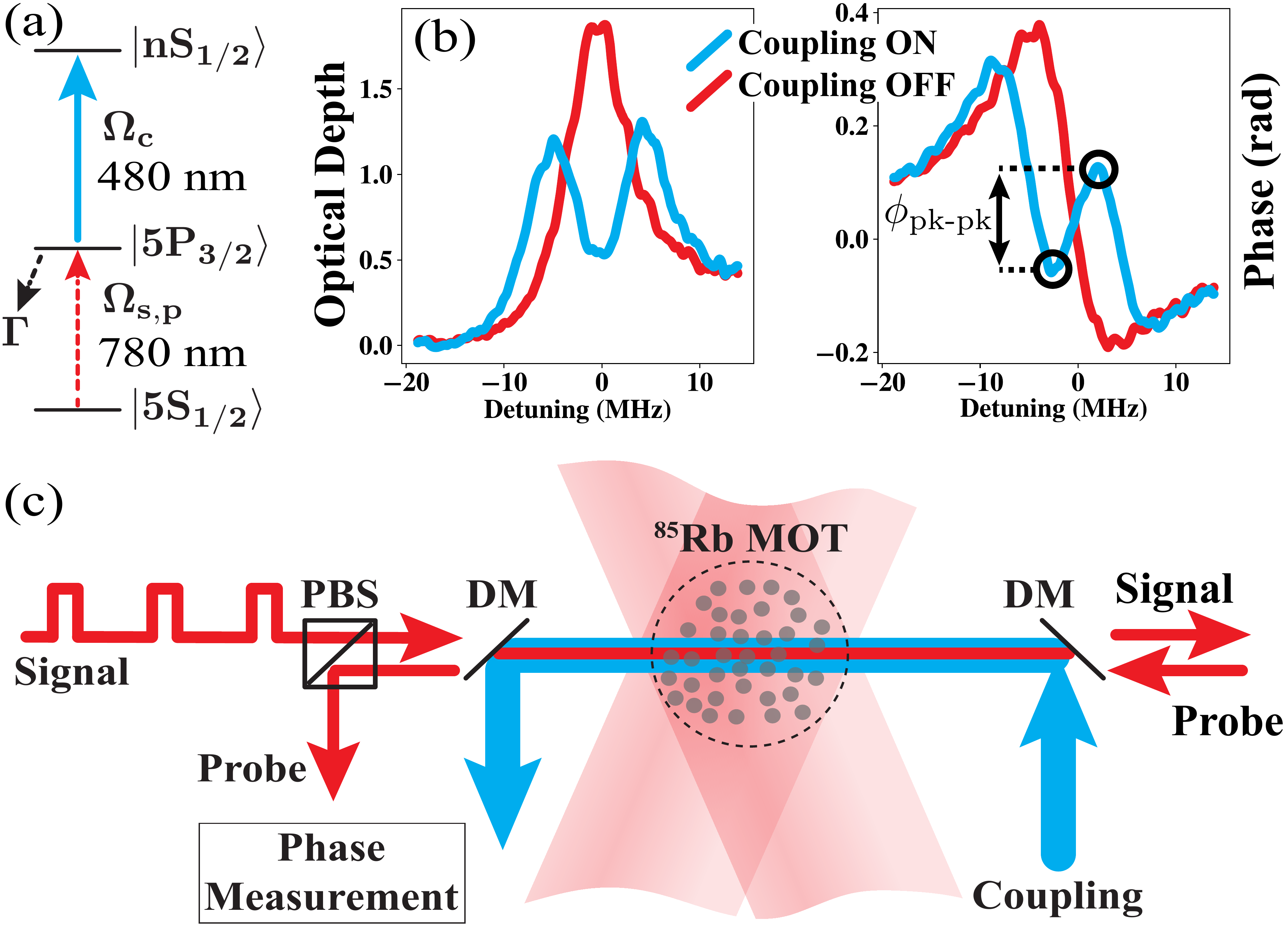}}
                \caption{(a) EIT level scheme.  (b) A typical spectroscopy scan taken during the spectroscopy stage of our duty cycle (n=58). From the spectrum we extract experimental parameters such as $\Omega_c$, peak OD with the coupling beam off and $\phi_\text{pk-pk}$. (c) Experimental setup. Dichroic mirrors (DM) are used to overlap the probe and signal with the coupling beam on top of the $^{85}$Rb magneto-optical trap (MOT). The probe and signal counterpropagate and are separated by polarizing beamsplitters (PBS). }
                \label{figure1} 
\end{figure}

\section{Theory}
In our scheme, the signal pulse ($\Omega_s$, see \hyperref[figure1]{Figure \ref{figure1}}a) propagates on resonance under EIT conditions created by a resonant coupling beam ($\Omega_c$). Inside the medium, signal photons excite Rydberg atoms, which interact via a van der Waals potential \cite{Singer2005} of the form 
\begin{equation}\label{VDWequation}
V(r) =  -\hbar C_6(n^*)/r^6,
\end{equation} 
where $C_6(n^*)$ characterizes the strength of the van der Waals potential, $n^*$ is the adjusted principal quantum number \cite{Wenhui2003}, and $r$ is the distance between two Rydberg atoms. Due to the interactions, once an atom is excited to its Rydberg state, atoms in its vicinity will experience a position-dependent shift in their Rydberg levels, leading to a shift in the index of refraction (see \hyperref[figure1]{Figure \ref{figure1}}b) seen by the probe ($\Omega_p$). Assuming perfect EIT (100\% transparency on resonance and no dephasing) the per-atom phase shift acquired by the probe is
\begin{equation}\label{atomphase}
    \phi_{X}(r) =  - \frac{\sigma}{\mathcal{A}} \cdot
    \frac{\Delta_{EIT} \: V(r)}{\Delta_{EIT}^2 + V(r)^2},
\end{equation}
where $\Delta_{EIT} = {\Omega_c^2}/{2 \Gamma}$ is the width of the EIT window and ${\sigma}/{\mathcal{A}}$ is the theoretical peak per-atom phase shift given by the probe transition's resonant cross-section ($\sigma$) divided by the area ($\mathcal{A}$) of the probe focus. From Eqn. \ref{atomphase}, we see that the phase shift acquired by the probe is maximized when $V(r)$ is comparable to $\Delta_{EIT}$, and goes to zero in the limit of either $V(r) \rightarrow 0$ or $V(r) \rightarrow \infty$. Considering a physical cloud of atoms with a small number of Rydberg excitations distributed randomly, neither the subset of atoms that are very far away from a Rydberg excitation, nor the subset that are very close impart phase shifts to the probe. Instead, it is the small subset of atoms whose distance from a Rydberg excitation falls within a shell, such that their Rydberg levels are shifted by approximately $\Delta_{EIT}$, which are responsible for shifting the phase of the probe. These are the atoms which are about $r_b = [C_6/\Delta_{EIT}]^{1/6}$ away from a Rydberg excitation, where $r_b$ is the blockade radius \cite{Parigi2012, Parigi, Gaetan2009, Abel2009}. The thickness of the shell, or rather, the range of distances where the interaction induces a shift comparable to $\Delta_{EIT}$ in the two-photon resonance, is determined (up to a numerical factor) by $r_b$. Remarkably, this means that the fraction of atoms in the cloud that reside in this shell, and therefore impart a significant phase shift to the probe, is proportional to $r_b^3$ (and therefore to $\sqrt{C_6/\Delta_{EIT}}$). This scaling is identical to the one observed in \cite{Parigi2012,grankin2014,Boddeda2016Experiment,Grankin2015}, however, the underlying model used to understand these experiments is markedly different. In \cite{Parigi2012,grankin2014,Boddeda2016Experiment,Grankin2015}, the effect of interest was dominated by the subset of atoms which reside within the volume $r_b^3$ and atoms farther than $r_b$ from a Rydberg excitation could be neglected.  Here, however, atoms inside and outside $r_b$ contribute equally and atoms deep inside $r_b$ do not contribute at all.

The total phase shift accumulated by the probe as it propagates through the medium can be found by summing up the contributions of each Rydberg excitation in the cloud
\begin{equation} \label{integral}
    \braket{\phi_{X}}= (\rho_\text{ryd} \mathcal{A} L) \int_0^{\infty}{\phi_{X}(r) \cdot (4 \pi r^2 \rho ) \cdot dr},
\end{equation}
where $\rho$ is the density of atoms, $\rho_\text{ryd}$ is the density of Rydberg excitations, and $L$ is the length of the medium \footnote{For simplicity we have assumed that the probability of Rydberg excitations is low enough to neglect any spatial correlations in the Rydberg density}. The exact result is
\begin{equation}\label{integralresult}
    \braket{\phi_{X}} =- \frac{\pi}{2\sqrt{2}} \cdot \textrm{OD} \cdot \left[{\frac{4}{3} \pi r_b^3 \: \rho_\text{ryd}}\right],
\end{equation}
where $\textrm{OD} = \rho \sigma L$.
At low signal power, the Rydberg density is linear in atom density and in the power ($\mathcal{P}_s \propto |\Omega_s|^2$) of the signal pulse:
\begin{equation} \label{density}
    \rho_\text{ryd} \approx \rho \cdot |\Omega_s|^2/|\Omega_c|^2 \propto \mathcal{P}_s. 
\end{equation}
Substituting this in to Eqn. \ref{integralresult}, and substituting $r_b^3 = \sqrt{C_6/\Delta_{EIT}}$, the total cross-phase shift is
\begin{equation}\label{finalresult}
    \braket{\phi_{X}} =- \frac{\sqrt{2} \pi^2 }{3} \cdot \textrm{OD} \cdot \sqrt{C_6/\Delta_{EIT}}\cdot  \frac{|\Omega_s|^2}{|\Omega_c|^2} \cdot \rho. 
\end{equation}
Eqn. \ref{finalresult} is the main result of our theoretical treatment: the phase shift written on the probe is proportional to the power of the signal beam. A distinguishing property of this Kerr nonlinearity is its dependence on $r_b^3$ which, in turn is proportional to $\sqrt{C_6(n^*)}$, making the phase proportional to $(n^*)^{5.5}$ when OD, $\Delta_{EIT}$, $\Omega_{c}$ are held constant. This scaling is dramatically different than what would be expected from a nonlinearity due to the AC Stark shift, which has no Rydberg level dependence, or superradiant cascade decay, which scales as $(n^*)^{2.5}$ \cite{Pritchard2011b}. Additionally, our Kerr-nonlinearity is cooperatively enhanced due to its quadratic dependence on atom density.

An important assumption made in the derivation of Eqn. \ref{finalresult} is ideal EIT. In our experiment, we observed between $75-50\%$ transparency on two-photon resonance (see \hyperref[figure1]{Figure \ref{figure1}}b) for several reasons including, frequency instability of the lasers used, partial overlap of the probe and coupling beams, and broadening and dephasing caused by Rydberg-Rydberg interactions. We can relax this assumption by replacing OD in Eqn. \ref{finalresult} with a measured quantity $\phi_\text{pk-pk}$, which we extract from $\phi(\Delta_p)$ the EIT spectrum (see \hyperref[figure1]{Figure \ref{figure1}}b). The main remaining approximation under which Eqn. \ref{finalresult} is valid is equivalent to requiring $\rho_\text{ryd} \ll [\frac{4}{3} \pi r_b^3]^{-1}$. This guarantees that the Rydberg density is spatially uniform and linear in the signal power. Maintaining a low Rydberg density requires low signal power ($\Omega_c \gg \Omega_s$), but also a low probe power ($\Omega_c \gg \Omega_p$). This second constraint is due to the fact that probe photons can also excite atoms into the Rydberg state. Although this process does not contribute to cross-phase modulation, and is therefore not included in our treatment, it does give rise to a second effect: probe photons create a background of Rydberg atoms. These can interact via van der Waals forces, or decay to nearby nP states and interact via stronger dipole-dipole forces \cite{Goldschmidt2017}, resulting in degraded electromagnetically induced transparency, reduced $\phi_\text{pk-pk}$, and smaller cross-phase shifts.

\section{Experiment}

The experimental scheme is shown in \hyperref[figure1]{Figure \ref{figure1}}c. We prepare a gas of about $10^8$ $^{85}$Rb  atoms in a magneto-optical trap (MOT) at a temperature of $60\pm10 \: \mu$K.  The MOT size is about $0.25 \text{mm}^3$ and the typical density is $\sim 3 \times 10^{10}  \: \text{atoms/cm}^3$. The atom duty cycle is 8.7ms long: it consists of 7.5ms of MOT trapping, molasses, and free expansion, followed by a 1.2ms measurement stage during which the phase and amplitude of the probe are measured via comparison to an off-resonant ``reference" beam using beat-note interferometry \cite{Dmochowski2012}. The probe is at 780 nm and is resonant with F=3 to F$'$=4. The reference is derived from the same master laser and is blue-detuned $2\pi \times 100$ MHz from the same transition. The signal counter-propagates (see \hyperref[figure1]{Figure \ref{figure1}}) with the probe and reference with orthogonal polarizations and is also resonant with  F=3 to F$'$=4. Setting the signal on resonance is particularly important in order to avoid the possibility of AC Stark shifts imparting an additional phase shift to the probe \cite{Feizpour2016d}. The typical resonant probe power was approximately 1 nW and the off-resonant ``reference" probe was about 10 nW. The coupling power varied from 10-200 mW depending on the Rydberg level and the signal power was varied between 10 pW to 100 nW. The probe and signal were focused to $20 \pm 4 \mu$m to ensure a constant waist over the length of the cloud ($L \approx 0.5$mm). The coupling beam was locked on a 5P$\rightarrow$nS resonance using EIT locking \cite{Abel2009} and was focused to $45 \pm 10 \mu$m in order to ensure a homogeneous coupling Rabi frequency across the transverse extent of the probe. The coupling Rabi frequency was typically $2 \pi \times (7 \pm 2)$ MHz, and the MOT OD was typically around 1-2. During the first 300\,$\mu$s of the measurement stage, spectroscopy scans were taken (\hyperref[figure1]{Figure \ref{figure1}}b depicts a typical spectroscopy scan for $n$=58. These scans are fit to theory curves to extract the peak optical depth without EIT, the optical depth with EIT, $\phi_\text{pk-pk}$, and the coupling Rabi frequency. For the remaining 900\,$\mu$s, the coupling and probe are left on and a signal pulse train consisting of 375 pulses, each 600 ns long and separated by 2.4\,$\mu$s, is turned on. In order to isolate the effect of the signal pulse on the phase of the probe from other slow drifts, the phase is measured before, during, and after each signal pulse. 
\begin{figure}[h]
        \centering
        \resizebox{8cm}{!}{
                \includegraphics[width=\textwidth]{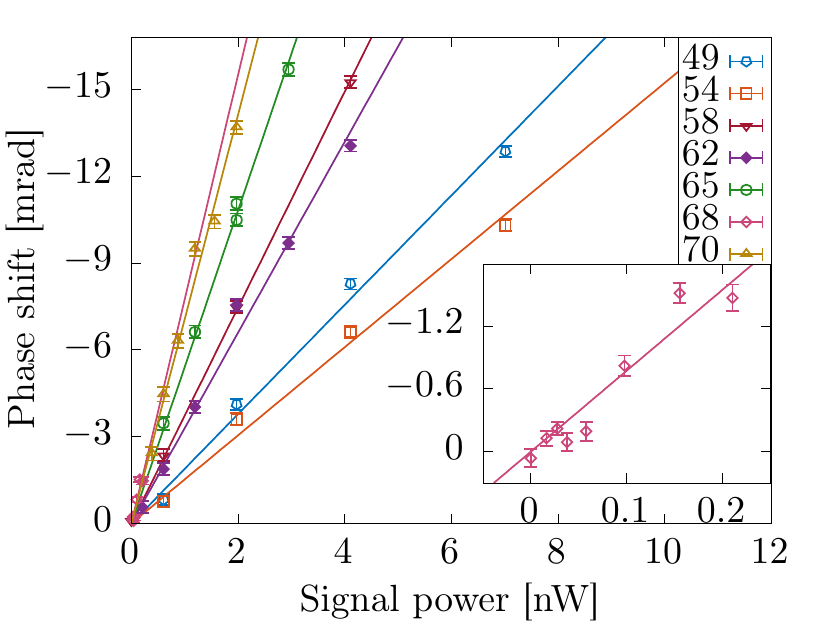}}
                \caption{The phase shift acquired by the probe is plotted as a function of signal power for seven different Rydberg levels. In the inset, n=68 is isolated to show that the phase shift is measured down to powers as low as $\sim 20$pW. Lines of best fit are plotted on top of the data for each Rydberg level. Error bars are the standard error of the mean (SEM) calculated from the data. The slope of the cross-phase shift generally increases with Rydberg level, with exceptions that we attribute to fluctuating $\phi_\text{pk-pk}$ }
                \label{PhaseShiftPerPhoton} 
\end{figure}
The measured shift in the phase of the probe is plotted in \hyperref[PhaseShiftPerPhoton]{Figure 2 \ref{PhaseShiftPerPhoton}} as a function of signal power for seven different S-state Rydberg levels (49, 54, 58, 62, 65, 68 and 70). The phase shift is linear up to 2-10 nW, depending on the Rydberg level, beyond which it saturates (data not shown). The slopes represent per-photon cross-phase shifts and generally increase with Rydberg level, with the largest slope at $n$=68, (due to a larger OD used for this Rydberg level) where approximately 8 mrad phase shifts were measured for 1 nW of signal power. From the probe EIT spectrum, we can infer that this corresponds to approximately 30 photons in the medium at a time (based off the interaction time ($\approx 8$ns) as opposed to the pulse duration). Additionally, linear cross-phase modulation was observed across nearly 3 orders of magnitude, persisting at signal powers as low as $\sim$20pW  as shown in the inset in \hyperref[PhaseShiftPerPhoton]{Figure 2\ref{PhaseShiftPerPhoton}} (corresponding to 47 photons in a 600 ns pulse, or $<1$ photons per interaction time).


\label{PhaseShiftPerPhoton} 

\begin{figure}[h]
        \centering
        \resizebox{8cm}{!}{
                \includegraphics[width=\textwidth]{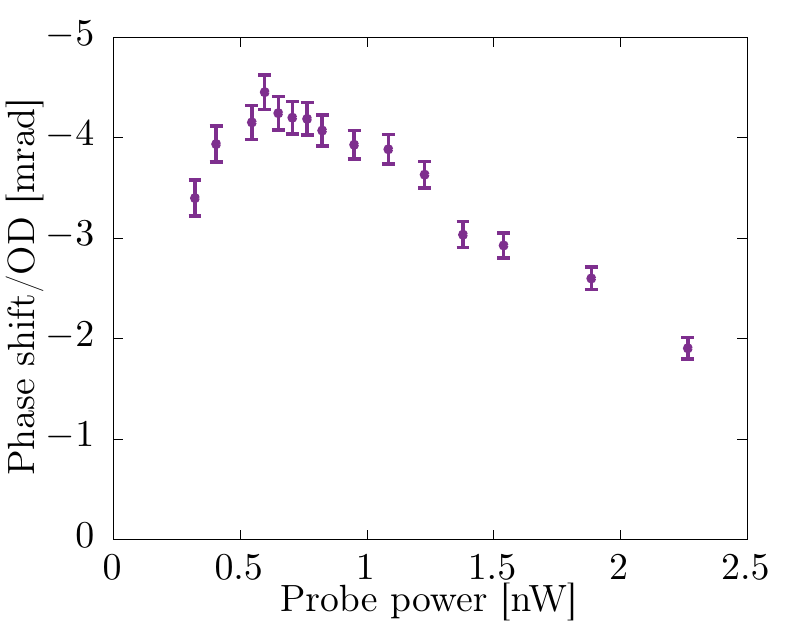}}
                \caption{The phase shift per OD acquired by the probe in the presence of 0.9 nW of signal power is plotted as a function of on-resonant probe power for $n=68$.  The coupling Rabi frequency is $2\pi \times (5.5 \pm 0.5)$ MHz. The OD varied between 1.1-2.2. }
                \label{ProbePowerDependence68S20180625}
\end{figure}

Next, the dependence of the cross-phase shift on probe power was studied. We observe reduced electromagnetically induced transparency as we increase the probe power, a problem which is exacerbated for higher Rydberg levels. In \hyperref[ProbePowerDependence68S20180625]{Figure \ref{ProbePowerDependence68S20180625}}, we vary the probe power and measure the cross-phase shift generated by $0.9$ nW of signal power for one of the higher Rydberg levels studied ($n=68$). During the experiment, the MOT OD drifted between approximately 1-2 and so, to compensate this source of variation, we study the cross-phase shifts divided by peak OD as measured during the spectroscopy stage for each probe power. At probe powers larger than approximately  1 nW, we see a significant decrease in the magnitude of the cross-phase shift. We attribute this to background Rydberg excitations created by the probe beam rather than the signal beam. The number of Rydberg excitations in the interaction region grows linearly with the total input photon number (signal plus probe) and once $\rho_\text{ryd}$ approaches $[\frac{4}{3} \pi r_b^3]^{-1}$, the interaction region is effectively saturated. Because the medium is blockaded, the signal beam cannot create additional Rydbergs and does not induce any cross-phase shift on the probe. Based on our focus size ($\mathcal{A}$), we expect to see $\rho_\text{ryd}$ become comparable to $[\frac{4}{3} \pi r_b^3]^{-1}$ when the total incident power reaches several nanoWatts, which is consistent with our observations in \hyperref[ProbePowerDependence68S20180625]{Figure \ref{ProbePowerDependence68S20180625}}. In the experiment, the probe power was typically kept at about 1 nW. Lower probe powers were not used due to increasing phase noise. To overcome this noise requires longer experimental runs which became technical challenging  due to experimental drift. (this may explain the unexpected drop on the leftmost points in \hyperref[ProbePowerDependence68S20180625]{Figure \ref{ProbePowerDependence68S20180625}})

\begin{figure}[h]
        \centering
        \resizebox{8cm}{!}{
                \includegraphics[width=\textwidth]{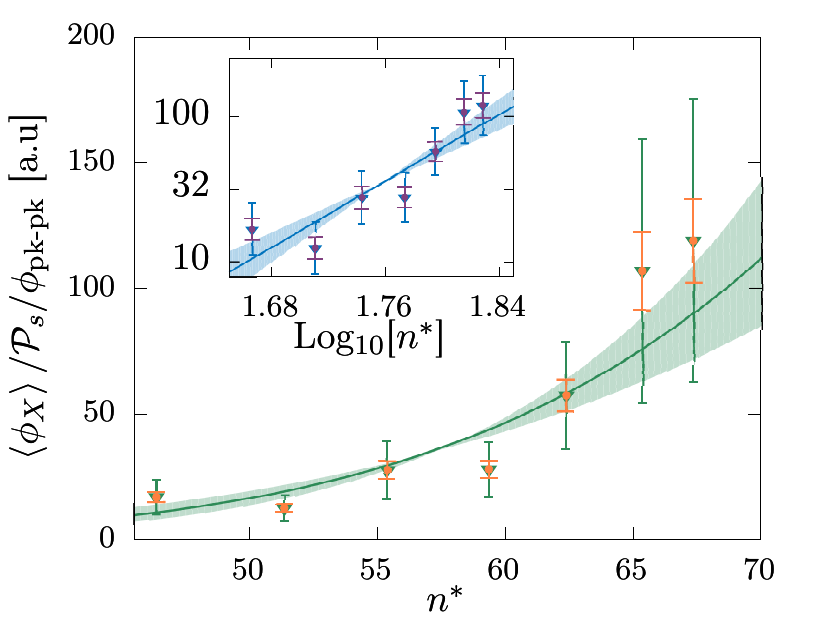}}
                \caption{$\Braket{\phi_{X}}/\mathcal{P}_s/\phi_\text{pk-pk}$ as a function of the adjusted Rydberg level. Here, $n^* = n - \delta$, where $\delta$ is the quantum defect and is approximately equal to 2.6. \cite{Wenhui2003}. The original error bars are shown in yellow (wider bars, circles) and the revised error bars in green (smaller bars, triangles). The data are fit (red (green) line) to a power law and the resulting fit has an exponent of $5.7 \pm 1.3$. The purple (blue) shaded region indicates fits within one sigma of this estimate (4.4-7). The inset shows the same data on a log scale.
                }
                \label{RydbergLevel} 
\end{figure}

We now turn to the Rydberg-level dependence of the cross-phase shift. The size of the Kerr nonlinearity, $\braket{\phi_{X}}/\mathcal{P}_s$, generally increased with Rydberg level. $\phi_\text{pk-pk}$ was also observed to vary strongly with Rydberg level, decreasing by roughly a factor of 2 between $n=49$ and $n=70$. This is likely due to background Rydbergs created by the probe, which saturate a fraction of the cloud that increases with $r_b^3$, contributing, among other things, to reduced transparency on two-photon resonance and smaller $\phi_\text{pk-pk}$. To eliminate this spurious-Rydberg-level dependence, we first divide $\braket{\phi_{X}}$ by the measured optical depth, replot it as a function of signal power and extract the slope for each Rydberg level using a linear fit. We then rescale the slopes by the ratio between $\phi_\text{pk-pk}$ and OD, which is measured for each Rydberg level. The rescaled slopes are then plotted as a function of Rydberg level in \hyperref[RydbergLevel]{Figure \ref{RydbergLevel}} and the same data is inset on a log-log plot, and fit to a line. The error bars are calculated from the statistical uncertainty of the linear fit (ranging between $2-10\%$), systemic uncertainty introduced by slow power drifts (estimated to be 10\%), and the uncertainty on $\phi_\text{pk-pk}$ (estimated to be $5-10\%$ based on noise in the measured spectra). The fit yields a power law with an exponent of $5.7\pm 0.4$, which is consistent with the predicted power law scaling, but the reduced $\chi^{2}$ (for our two parameter fit to seven data points) is $11$. We conclude from this large reduced $\chi^{2}$ that we have underestimated the degree to which inaccuracies in our rescaling technique changed between runs with different principal quantum numbers by a factor on the order of $\sqrt{11}$. We use the revised estimate to extract a conservative uncertainty on the fit parameter, now $5.7\pm 1.3$. This result rules out the AC Stark shift as an explanation for our nonlinearity, and leads us to conclude that van der Waals-based interactions are the best explanation for the observed cross-Kerr nonlinearity.


\section{Discussion}
Finally, we compare the size of our Rydberg-based cross-Kerr nonlinearity to those observed in other systems. The maximum cross-phase shift observed was 8 mrad/nW (for $n=68$). This corresponds to an effective value of Re$\left[\chi^{(3)}\right] \sim 1\times 10^{-8} \textrm{m}^2/\textrm{V}^2$. This is much larger than the Re$\left[\chi^{(3)}\right]$ available in conventional materials like fused silica \cite{Boyd3rd}, and similar in magnitude to the Re$\left[\chi^{(3)}\right]$ reported in \cite{Parigi2012} (see table), in which Parigi \textit{et. al.} studied a cavity-enhanced self-Kerr effect based on off-resonant Rydberg EIT. It is also worth noting that while larger nonlinearities have been observed in interacting Rydberg gases, for example \cite{Pritchard2011a}, they were dissipative (the measured $\chi^{(3)}$ was imaginary), which makes them unsuitable for applications to quantum state generation and quantum non-demolition measurement of photon number. Going beyond Rydberg-based Kerr nonlinearities, our Re$\left[\chi^{(3)}\right]$ is 5000 times larger than the nonlinearity used by Venkataraman et. al. \cite{Venkataraman2013} to observe cross-phase modulation in a hollow core fibre loaded with ultracold Rubidium atoms. In fact, our Re$\left[\chi^{(3)}\right]$ is only 50 times smaller than the record Re$\left[\chi^{(3)}\right]$ observed in slow-light experiments at BEC densities despite the fact that our experiment was performed in a simple magneto-optical trap with 100 times lower density \cite{Hau1999a}. Finally, our measured Re$\left[\chi^{(3)}\right]$ is 5 times larger than the Re$\left[\chi^{(3)}\right]$ measured by Feizpour et. al. based on EIT and AC Stark shifts (N-scheme) \cite{Feizpour2015} . 

For many of the applications of strong single-photon-level nonlinearities a more important figure of merit than Re$\left[\chi^{(3)}\right]$ is $\phi_0$, the phase shift per photon. Several of the experiments discussed here observed phase shifts per photon that ranged from 13 $\mu$rad/photon \cite{Feizpour2015} to 300 $\mu$rad/photon \cite{Venkataraman2013}. Because our experiment was performed with long signal pulses we did not \textit{directly} measure the phase shift per signal photon. We did, however, measure $\phi(\Delta_p)$ during our spectroscopy stage and can therefore infer the group velocity and the interaction time, allowing us to infer the average number of photons in the atomic medium at one time for a long signal pulse with a peak power of 1 nW. In this way, we can indirectly estimate the \textit{per-photon} phase shift. For n=68, the group delay was approximately 8 ns and the corresponding per-photon phase shift was $\sim -250 \: \mu$rad/photon.

\begin{center}
 \begin{tabular}{||c | c | c | c ||} 
 \hline
 Description  & $\chi^{(3)}$ (m$^2/$V$^2$) & $|\phi_0|$ ($\mu$rad/ph.)   & Ref. \\ [0.5ex] \hline

 \hline
 Fused Silica & $2.5 \times 10^{-22}$ & & \cite{Boyd3rd}\\
 \hline
    EIT, BEC & $5 \times 10^{-7} $ & & \cite{Hau1999a} \\
 \hline
  Ryd., S-Kerr & $ i \cdot 5 \times 10^{-7}$ & & \cite{Pritchard2011a} \\
 \hline
 Ryd., X-Kerr  & $1 \times 10^{-8}$  & 250$^{\star}$ & \smiley{} \\ 
 \hline
 Ryd., S-Kerr & $5 \times 10^{-9} $ & & \cite{Parigi2012} \\
 \hline
  N-scheme MOT & $2 \times 10^{-9}$ & 13 & \cite{Feizpour2015} \\
 \hline 
 N-scheme HCF* & $1 \times 10^{-12}$ &300 & \cite{Venkataraman2013} \\
\hline

\end{tabular}\par
\bigskip
\label{tab:table-name}
Table 1: Comparison of selected Kerr nonlinearities.  (* hollow core fibre, $\star \text{ not directly measured,} \: \: \smiley \:\text{this work}$).
\end{center}

\section{Conclusion}
In conclusion, we observed strong cross-phase modulation at low light levels, generated by a resonant cross-Kerr nonlinearity based on Ryberg-Ryberg interactions and EIT. We directly measured cross phase shifts and estimate the Re$\left[\chi^{(3)}\right]$ to be $10^{-8} \: \textrm{m}^2/\textrm{V}^2$. We varied the Rydberg level and observed a scaling with the principal quantum number consistent with the $(n^*)^{5.5}$ expected for a van der Waals-based nonlinearity. This is the first experimental demonstration of a cross-Kerr nonlinearity between two beams based on Rydberg interactions. Future experiments will explore ways to reduce self-phase interactions, which, combined with higher density and higher Rydberg levels, should enable even larger single-photon level dispersive nonlinearities with important applications in quantum optics and quantum information.

\section{acknowledgements}
This research was supported by NSERC, CIFAR, and the Fetzer Franklin Fund of the John E. Fetzer Memorial Trust. The authors are grateful to Hudson Pimenta for valuable discussions, and Alan Stummer for equipment design.

\bibliography{Kerr}

\begin{thebibliography}{42}%
\makeatletter
\providecommand \@ifxundefined [1]{%
 \@ifx{#1\undefined}
}%
\providecommand \@ifnum [1]{%
 \ifnum #1\expandafter \@firstoftwo
 \else \expandafter \@secondoftwo
 \fi
}%
\providecommand \@ifx [1]{%
 \ifx #1\expandafter \@firstoftwo
 \else \expandafter \@secondoftwo
 \fi
}%
\providecommand \natexlab [1]{#1}%
\providecommand \enquote  [1]{``#1''}%
\providecommand \bibnamefont  [1]{#1}%
\providecommand \bibfnamefont [1]{#1}%
\providecommand \citenamefont [1]{#1}%
\providecommand \href@noop [0]{\@secondoftwo}%
\providecommand \href [0]{\begingroup \@sanitize@url \@href}%
\providecommand \@href[1]{\@@startlink{#1}\@@href}%
\providecommand \@@href[1]{\endgroup#1\@@endlink}%
\providecommand \@sanitize@url [0]{\catcode `\\12\catcode `\$12\catcode
  `\&12\catcode `\#12\catcode `\^12\catcode `\_12\catcode `\%12\relax}%
\providecommand \@@startlink[1]{}%
\providecommand \@@endlink[0]{}%
\providecommand \url  [0]{\begingroup\@sanitize@url \@url }%
\providecommand \@url [1]{\endgroup\@href {#1}{\urlprefix }}%
\providecommand \urlprefix  [0]{URL }%
\providecommand \Eprint [0]{\href }%
\providecommand \doibase [0]{http://dx.doi.org/}%
\providecommand \selectlanguage [0]{\@gobble}%
\providecommand \bibinfo  [0]{\@secondoftwo}%
\providecommand \bibfield  [0]{\@secondoftwo}%
\providecommand \translation [1]{[#1]}%
\providecommand \BibitemOpen [0]{}%
\providecommand \bibitemStop [0]{}%
\providecommand \bibitemNoStop [0]{.\EOS\space}%
\providecommand \EOS [0]{\spacefactor3000\relax}%
\providecommand \BibitemShut  [1]{\csname bibitem#1\endcsname}%
\let\auto@bib@innerbib\@empty
\bibitem [{\citenamefont {Tiarks}\ \emph {et~al.}(2018)\citenamefont {Tiarks},
  \citenamefont {Schmidt-Eberle}, \citenamefont {Stolz}, \citenamefont
  {Rempe},\ and\ \citenamefont {D{\"u}rr}}]{Tiarks2018}%
  \BibitemOpen
  \bibfield  {author} {\bibinfo {author} {\bibfnamefont {D.}~\bibnamefont
  {Tiarks}}, \bibinfo {author} {\bibfnamefont {S.}~\bibnamefont
  {Schmidt-Eberle}}, \bibinfo {author} {\bibfnamefont {T.}~\bibnamefont
  {Stolz}}, \bibinfo {author} {\bibfnamefont {G.}~\bibnamefont {Rempe}}, \ and\
  \bibinfo {author} {\bibfnamefont {S.}~\bibnamefont {D{\"u}rr}},\ }\href
  {\doibase 10.1038/s41567-018-0313-7} {\bibfield  {journal} {\bibinfo
  {journal} {Nature Physics}\ } (\bibinfo {year} {2018}),\
  10.1038/s41567-018-0313-7}\BibitemShut {NoStop}%
\bibitem [{\citenamefont {Baur}\ \emph {et~al.}(2014)\citenamefont {Baur},
  \citenamefont {Tiarks}, \citenamefont {Rempe},\ and\ \citenamefont
  {D\"urr}}]{Baur2014a}%
  \BibitemOpen
  \bibfield  {author} {\bibinfo {author} {\bibfnamefont {S.}~\bibnamefont
  {Baur}}, \bibinfo {author} {\bibfnamefont {D.}~\bibnamefont {Tiarks}},
  \bibinfo {author} {\bibfnamefont {G.}~\bibnamefont {Rempe}}, \ and\ \bibinfo
  {author} {\bibfnamefont {S.}~\bibnamefont {D\"urr}},\ }\href {\doibase
  10.1103/PhysRevLett.112.073901} {\bibfield  {journal} {\bibinfo  {journal}
  {Phys. Rev. Lett.}\ }\textbf {\bibinfo {volume} {112}},\ \bibinfo {pages}
  {073901} (\bibinfo {year} {2014})}\BibitemShut {NoStop}%
\bibitem [{\citenamefont {M\"uller}\ \emph {et~al.}(2009)\citenamefont
  {M\"uller}, \citenamefont {Lesanovsky}, \citenamefont {Weimer}, \citenamefont
  {B\"uchler},\ and\ \citenamefont {Zoller}}]{Muller2009}%
  \BibitemOpen
  \bibfield  {author} {\bibinfo {author} {\bibfnamefont {M.}~\bibnamefont
  {M\"uller}}, \bibinfo {author} {\bibfnamefont {I.}~\bibnamefont
  {Lesanovsky}}, \bibinfo {author} {\bibfnamefont {H.}~\bibnamefont {Weimer}},
  \bibinfo {author} {\bibfnamefont {H.~P.}\ \bibnamefont {B\"uchler}}, \ and\
  \bibinfo {author} {\bibfnamefont {P.}~\bibnamefont {Zoller}},\ }\href
  {\doibase 10.1103/PhysRevLett.102.170502} {\bibfield  {journal} {\bibinfo
  {journal} {Phys. Rev. Lett.}\ }\textbf {\bibinfo {volume} {102}},\ \bibinfo
  {pages} {170502} (\bibinfo {year} {2009})}\BibitemShut {NoStop}%
\bibitem [{\citenamefont {Keating}\ \emph {et~al.}(2015)\citenamefont
  {Keating}, \citenamefont {Cook}, \citenamefont {Hankin}, \citenamefont {Jau},
  \citenamefont {Biedermann},\ and\ \citenamefont {Deutsch}}]{Keating}%
  \BibitemOpen
  \bibfield  {author} {\bibinfo {author} {\bibfnamefont {T.}~\bibnamefont
  {Keating}}, \bibinfo {author} {\bibfnamefont {R.~L.}\ \bibnamefont {Cook}},
  \bibinfo {author} {\bibfnamefont {A.~M.}\ \bibnamefont {Hankin}}, \bibinfo
  {author} {\bibfnamefont {Y.-Y.}\ \bibnamefont {Jau}}, \bibinfo {author}
  {\bibfnamefont {G.~W.}\ \bibnamefont {Biedermann}}, \ and\ \bibinfo {author}
  {\bibfnamefont {I.~H.}\ \bibnamefont {Deutsch}},\ }\href {\doibase
  10.1103/PhysRevA.91.012337} {\bibfield  {journal} {\bibinfo  {journal} {Phys.
  Rev. A}\ }\textbf {\bibinfo {volume} {91}},\ \bibinfo {pages} {012337}
  (\bibinfo {year} {2015})}\BibitemShut {NoStop}%
\bibitem [{\citenamefont {Distante}\ \emph {et~al.}(2016)\citenamefont
  {Distante}, \citenamefont {Padr\'on-Brito}, \citenamefont {Cristiani},
  \citenamefont {Paredes-Barato},\ and\ \citenamefont
  {de~Riedmatten}}]{Distante2016}%
  \BibitemOpen
  \bibfield  {author} {\bibinfo {author} {\bibfnamefont {E.}~\bibnamefont
  {Distante}}, \bibinfo {author} {\bibfnamefont {A.}~\bibnamefont
  {Padr\'on-Brito}}, \bibinfo {author} {\bibfnamefont {M.}~\bibnamefont
  {Cristiani}}, \bibinfo {author} {\bibfnamefont {D.}~\bibnamefont
  {Paredes-Barato}}, \ and\ \bibinfo {author} {\bibfnamefont {H.}~\bibnamefont
  {de~Riedmatten}},\ }\href {\doibase 10.1103/PhysRevLett.117.113001}
  {\bibfield  {journal} {\bibinfo  {journal} {Phys. Rev. Lett.}\ }\textbf
  {\bibinfo {volume} {117}},\ \bibinfo {pages} {113001} (\bibinfo {year}
  {2016})}\BibitemShut {NoStop}%
\bibitem [{\citenamefont {Maxwell}\ \emph {et~al.}(2013)\citenamefont
  {Maxwell}, \citenamefont {Szwer}, \citenamefont {Paredes-Barato},
  \citenamefont {Busche}, \citenamefont {Pritchard}, \citenamefont {Gauguet},
  \citenamefont {Weatherill}, \citenamefont {Jones},\ and\ \citenamefont
  {Adams}}]{Maxwell2013}%
  \BibitemOpen
  \bibfield  {author} {\bibinfo {author} {\bibfnamefont {D.}~\bibnamefont
  {Maxwell}}, \bibinfo {author} {\bibfnamefont {D.~J.}\ \bibnamefont {Szwer}},
  \bibinfo {author} {\bibfnamefont {D.}~\bibnamefont {Paredes-Barato}},
  \bibinfo {author} {\bibfnamefont {H.}~\bibnamefont {Busche}}, \bibinfo
  {author} {\bibfnamefont {J.~D.}\ \bibnamefont {Pritchard}}, \bibinfo {author}
  {\bibfnamefont {A.}~\bibnamefont {Gauguet}}, \bibinfo {author} {\bibfnamefont
  {K.~J.}\ \bibnamefont {Weatherill}}, \bibinfo {author} {\bibfnamefont
  {M.~P.~A.}\ \bibnamefont {Jones}}, \ and\ \bibinfo {author} {\bibfnamefont
  {C.~S.}\ \bibnamefont {Adams}},\ }\href {\doibase
  10.1103/PhysRevLett.110.103001} {\bibfield  {journal} {\bibinfo  {journal}
  {Phys. Rev. Lett.}\ }\textbf {\bibinfo {volume} {110}},\ \bibinfo {pages}
  {103001} (\bibinfo {year} {2013})}\BibitemShut {NoStop}%
\bibitem [{\citenamefont {Dudin}\ and\ \citenamefont
  {Kuzmich}(2012)}]{Dudin2012a}%
  \BibitemOpen
  \bibfield  {author} {\bibinfo {author} {\bibfnamefont {Y.~O.}\ \bibnamefont
  {Dudin}}\ and\ \bibinfo {author} {\bibfnamefont {A.}~\bibnamefont
  {Kuzmich}},\ }\href {\doibase 10.1126/science.1217901} {\bibfield  {journal}
  {\bibinfo  {journal} {Science}\ }\textbf {\bibinfo {volume} {336}},\ \bibinfo
  {pages} {887} (\bibinfo {year} {2012})}\BibitemShut {NoStop}%
\bibitem [{\citenamefont {Ripka}\ \emph {et~al.}(2018)\citenamefont {Ripka},
  \citenamefont {K{\"u}bler}, \citenamefont {L{\"o}w},\ and\ \citenamefont
  {Pfau}}]{Ripka446}%
  \BibitemOpen
  \bibfield  {author} {\bibinfo {author} {\bibfnamefont {F.}~\bibnamefont
  {Ripka}}, \bibinfo {author} {\bibfnamefont {H.}~\bibnamefont {K{\"u}bler}},
  \bibinfo {author} {\bibfnamefont {R.}~\bibnamefont {L{\"o}w}}, \ and\
  \bibinfo {author} {\bibfnamefont {T.}~\bibnamefont {Pfau}},\ }\href {\doibase
  10.1126/science.aau1949} {\bibfield  {journal} {\bibinfo  {journal}
  {Science}\ }\textbf {\bibinfo {volume} {362}},\ \bibinfo {pages} {446}
  (\bibinfo {year} {2018})}\BibitemShut {NoStop}%
\bibitem [{\citenamefont {Firstenberg}\ \emph {et~al.}(2013)\citenamefont
  {Firstenberg}, \citenamefont {Peyronel}, \citenamefont {Liang}, \citenamefont
  {Gorshkov}, \citenamefont {Lukin},\ and\ \citenamefont
  {Vuletic}}]{Firstenberg2013}%
  \BibitemOpen
  \bibfield  {author} {\bibinfo {author} {\bibfnamefont {O.}~\bibnamefont
  {Firstenberg}}, \bibinfo {author} {\bibfnamefont {T.}~\bibnamefont
  {Peyronel}}, \bibinfo {author} {\bibfnamefont {Q.-Y.}\ \bibnamefont {Liang}},
  \bibinfo {author} {\bibfnamefont {A.~V.}\ \bibnamefont {Gorshkov}}, \bibinfo
  {author} {\bibfnamefont {M.~D.}\ \bibnamefont {Lukin}}, \ and\ \bibinfo
  {author} {\bibfnamefont {V.}~\bibnamefont {Vuletic}},\ }\href
  {https://doi.org/10.1038/nature12512} {\bibfield  {journal} {\bibinfo
  {journal} {Nature}\ }\textbf {\bibinfo {volume} {502}},\ \bibinfo {pages} {71
  EP } (\bibinfo {year} {2013})}\BibitemShut {NoStop}%
\bibitem [{\citenamefont {Peyronel}\ \emph {et~al.}(2012)\citenamefont
  {Peyronel}, \citenamefont {Firstenberg}, \citenamefont {Liang}, \citenamefont
  {Hofferberth}, \citenamefont {Gorshkov}, \citenamefont {Pohl}, \citenamefont
  {Lukin},\ and\ \citenamefont {Vuletic}}]{Peyronel2012c}%
  \BibitemOpen
  \bibfield  {author} {\bibinfo {author} {\bibfnamefont {T.}~\bibnamefont
  {Peyronel}}, \bibinfo {author} {\bibfnamefont {O.}~\bibnamefont
  {Firstenberg}}, \bibinfo {author} {\bibfnamefont {Q.-Y.}\ \bibnamefont
  {Liang}}, \bibinfo {author} {\bibfnamefont {S.}~\bibnamefont {Hofferberth}},
  \bibinfo {author} {\bibfnamefont {A.~V.}\ \bibnamefont {Gorshkov}}, \bibinfo
  {author} {\bibfnamefont {T.}~\bibnamefont {Pohl}}, \bibinfo {author}
  {\bibfnamefont {M.~D.}\ \bibnamefont {Lukin}}, \ and\ \bibinfo {author}
  {\bibfnamefont {V.}~\bibnamefont {Vuletic}},\ }\href
  {https://doi.org/10.1038/nature11361} {\bibfield  {journal} {\bibinfo
  {journal} {Nature}\ }\textbf {\bibinfo {volume} {488}},\ \bibinfo {pages} {57
  EP } (\bibinfo {year} {2012})}\BibitemShut {NoStop}%
\bibitem [{\citenamefont {Thompson}\ \emph {et~al.}(2017)\citenamefont
  {Thompson}, \citenamefont {Nicholson}, \citenamefont {Liang}, \citenamefont
  {Cantu}, \citenamefont {Venkatramani}, \citenamefont {Choi}, \citenamefont
  {Fedorov}, \citenamefont {Viscor}, \citenamefont {Pohl}, \citenamefont
  {Lukin},\ and\ \citenamefont {Vuletic}}]{Thompson2017}%
  \BibitemOpen
  \bibfield  {author} {\bibinfo {author} {\bibfnamefont {J.~D.}\ \bibnamefont
  {Thompson}}, \bibinfo {author} {\bibfnamefont {T.~L.}\ \bibnamefont
  {Nicholson}}, \bibinfo {author} {\bibfnamefont {Q.-Y.}\ \bibnamefont
  {Liang}}, \bibinfo {author} {\bibfnamefont {S.~H.}\ \bibnamefont {Cantu}},
  \bibinfo {author} {\bibfnamefont {A.~V.}\ \bibnamefont {Venkatramani}},
  \bibinfo {author} {\bibfnamefont {S.}~\bibnamefont {Choi}}, \bibinfo {author}
  {\bibfnamefont {I.~A.}\ \bibnamefont {Fedorov}}, \bibinfo {author}
  {\bibfnamefont {D.}~\bibnamefont {Viscor}}, \bibinfo {author} {\bibfnamefont
  {T.}~\bibnamefont {Pohl}}, \bibinfo {author} {\bibfnamefont {M.~D.}\
  \bibnamefont {Lukin}}, \ and\ \bibinfo {author} {\bibfnamefont
  {V.}~\bibnamefont {Vuletic}},\ }\href {https://doi.org/10.1038/nature20823}
  {\bibfield  {journal} {\bibinfo  {journal} {Nature}\ }\textbf {\bibinfo
  {volume} {542}},\ \bibinfo {pages} {206 EP } (\bibinfo {year}
  {2017})}\BibitemShut {NoStop}%
\bibitem [{\citenamefont {Liang}\ \emph {et~al.}(2018)\citenamefont {Liang},
  \citenamefont {Venkatramani}, \citenamefont {Cantu}, \citenamefont
  {Nicholson}, \citenamefont {Gullans}, \citenamefont {Gorshkov}, \citenamefont
  {Thompson}, \citenamefont {Chin}, \citenamefont {Lukin},\ and\ \citenamefont
  {Vuleti{\'c}}}]{Liang2018}%
  \BibitemOpen
  \bibfield  {author} {\bibinfo {author} {\bibfnamefont {Q.-Y.}\ \bibnamefont
  {Liang}}, \bibinfo {author} {\bibfnamefont {A.~V.}\ \bibnamefont
  {Venkatramani}}, \bibinfo {author} {\bibfnamefont {S.~H.}\ \bibnamefont
  {Cantu}}, \bibinfo {author} {\bibfnamefont {T.~L.}\ \bibnamefont
  {Nicholson}}, \bibinfo {author} {\bibfnamefont {M.~J.}\ \bibnamefont
  {Gullans}}, \bibinfo {author} {\bibfnamefont {A.~V.}\ \bibnamefont
  {Gorshkov}}, \bibinfo {author} {\bibfnamefont {J.~D.}\ \bibnamefont
  {Thompson}}, \bibinfo {author} {\bibfnamefont {C.}~\bibnamefont {Chin}},
  \bibinfo {author} {\bibfnamefont {M.~D.}\ \bibnamefont {Lukin}}, \ and\
  \bibinfo {author} {\bibfnamefont {V.}~\bibnamefont {Vuleti{\'c}}},\ }\href
  {\doibase 10.1126/science.aao7293} {\bibfield  {journal} {\bibinfo  {journal}
  {Science}\ }\textbf {\bibinfo {volume} {359}},\ \bibinfo {pages} {783}
  (\bibinfo {year} {2018})}\BibitemShut {NoStop}%
\bibitem [{\citenamefont {Nemoto}\ and\ \citenamefont
  {Munro}(2004)}]{NemotoMunro2004}%
  \BibitemOpen
  \bibfield  {author} {\bibinfo {author} {\bibfnamefont {K.}~\bibnamefont
  {Nemoto}}\ and\ \bibinfo {author} {\bibfnamefont {W.~J.}\ \bibnamefont
  {Munro}},\ }\href {\doibase 10.1103/PhysRevLett.93.250502} {\bibfield
  {journal} {\bibinfo  {journal} {Phys. Rev. Lett.}\ }\textbf {\bibinfo
  {volume} {93}},\ \bibinfo {pages} {250502} (\bibinfo {year}
  {2004})}\BibitemShut {NoStop}%
\bibitem [{\citenamefont {Munro}\ \emph {et~al.}(2005)\citenamefont {Munro},
  \citenamefont {Nemoto},\ and\ \citenamefont {Spiller}}]{MunroNemoto2005}%
  \BibitemOpen
  \bibfield  {author} {\bibinfo {author} {\bibfnamefont {W.~J.}\ \bibnamefont
  {Munro}}, \bibinfo {author} {\bibfnamefont {K.}~\bibnamefont {Nemoto}}, \
  and\ \bibinfo {author} {\bibfnamefont {T.~P.}\ \bibnamefont {Spiller}},\
  }\href {http://stacks.iop.org/1367-2630/7/i=1/a=137} {\bibfield  {journal}
  {\bibinfo  {journal} {New Journal of Physics}\ }\textbf {\bibinfo {volume}
  {7}},\ \bibinfo {pages} {137} (\bibinfo {year} {2005})}\BibitemShut {NoStop}%
\bibitem [{\citenamefont {Imoto}\ \emph {et~al.}(1985)\citenamefont {Imoto},
  \citenamefont {Haus},\ and\ \citenamefont {Yamamoto}}]{Imoto1985}%
  \BibitemOpen
  \bibfield  {author} {\bibinfo {author} {\bibfnamefont {N.}~\bibnamefont
  {Imoto}}, \bibinfo {author} {\bibfnamefont {H.~A.}\ \bibnamefont {Haus}}, \
  and\ \bibinfo {author} {\bibfnamefont {Y.}~\bibnamefont {Yamamoto}},\ }\href
  {\doibase 10.1103/PhysRevA.32.2287} {\bibfield  {journal} {\bibinfo
  {journal} {Physical Review A}\ }\textbf {\bibinfo {volume} {32}},\ \bibinfo
  {pages} {2287} (\bibinfo {year} {1985})}\BibitemShut {NoStop}%
\bibitem [{\citenamefont {Schmid}\ \emph {et~al.}(2017)\citenamefont {Schmid},
  \citenamefont {Marshall},\ and\ \citenamefont {James}}]{Schmid2017}%
  \BibitemOpen
  \bibfield  {author} {\bibinfo {author} {\bibfnamefont {D.}~\bibnamefont
  {Schmid}}, \bibinfo {author} {\bibfnamefont {K.}~\bibnamefont {Marshall}}, \
  and\ \bibinfo {author} {\bibfnamefont {D.~F.~V.}\ \bibnamefont {James}},\
  }\href {\doibase 10.1080/09500340.2017.1357852} {\bibfield  {journal}
  {\bibinfo  {journal} {Journal of Modern Optics}\ }\textbf {\bibinfo {volume}
  {64}},\ \bibinfo {pages} {2306} (\bibinfo {year} {2017})}\BibitemShut
  {NoStop}%
\bibitem [{\citenamefont {Bai}\ and\ \citenamefont {Huang}(2016)}]{Bai2008}%
  \BibitemOpen
  \bibfield  {author} {\bibinfo {author} {\bibfnamefont {Z.}~\bibnamefont
  {Bai}}\ and\ \bibinfo {author} {\bibfnamefont {G.}~\bibnamefont {Huang}},\
  }\href {\doibase 10.1364/OE.24.004442} {\bibfield  {journal} {\bibinfo
  {journal} {Opt. Express}\ }\textbf {\bibinfo {volume} {24}},\ \bibinfo
  {pages} {4442} (\bibinfo {year} {2016})}\BibitemShut {NoStop}%
\bibitem [{\citenamefont {Ates}\ \emph {et~al.}(2011)\citenamefont {Ates},
  \citenamefont {Sevin\ifmmode~\mbox{\c{c}}\else \c{c}\fi{}li},\ and\
  \citenamefont {Pohl}}]{Pohl2011a}%
  \BibitemOpen
  \bibfield  {author} {\bibinfo {author} {\bibfnamefont {C.}~\bibnamefont
  {Ates}}, \bibinfo {author} {\bibfnamefont {S.}~\bibnamefont
  {Sevin\ifmmode~\mbox{\c{c}}\else \c{c}\fi{}li}}, \ and\ \bibinfo {author}
  {\bibfnamefont {T.}~\bibnamefont {Pohl}},\ }\href {\doibase
  10.1103/PhysRevA.83.041802} {\bibfield  {journal} {\bibinfo  {journal} {Phys.
  Rev. A}\ }\textbf {\bibinfo {volume} {83}},\ \bibinfo {pages} {041802}
  (\bibinfo {year} {2011})}\BibitemShut {NoStop}%
\bibitem [{\citenamefont {Sevin\ifmmode~\mbox{\c{c}}\else \c{c}\fi{}li}\ \emph
  {et~al.}(2011)\citenamefont {Sevin\ifmmode~\mbox{\c{c}}\else \c{c}\fi{}li},
  \citenamefont {Henkel}, \citenamefont {Ates},\ and\ \citenamefont
  {Pohl}}]{Pohl2011b}%
  \BibitemOpen
  \bibfield  {author} {\bibinfo {author} {\bibfnamefont {S.}~\bibnamefont
  {Sevin\ifmmode~\mbox{\c{c}}\else \c{c}\fi{}li}}, \bibinfo {author}
  {\bibfnamefont {N.}~\bibnamefont {Henkel}}, \bibinfo {author} {\bibfnamefont
  {C.}~\bibnamefont {Ates}}, \ and\ \bibinfo {author} {\bibfnamefont
  {T.}~\bibnamefont {Pohl}},\ }\href {\doibase 10.1103/PhysRevLett.107.153001}
  {\bibfield  {journal} {\bibinfo  {journal} {Phys. Rev. Lett.}\ }\textbf
  {\bibinfo {volume} {107}},\ \bibinfo {pages} {153001} (\bibinfo {year}
  {2011})}\BibitemShut {NoStop}%
\bibitem [{\citenamefont {Stanojevic}\ \emph {et~al.}(2013)\citenamefont
  {Stanojevic}, \citenamefont {Parigi}, \citenamefont {Bimbard}, \citenamefont
  {Ourjoumtsev},\ and\ \citenamefont {Grangier}}]{Grangier2013}%
  \BibitemOpen
  \bibfield  {author} {\bibinfo {author} {\bibfnamefont {J.}~\bibnamefont
  {Stanojevic}}, \bibinfo {author} {\bibfnamefont {V.}~\bibnamefont {Parigi}},
  \bibinfo {author} {\bibfnamefont {E.}~\bibnamefont {Bimbard}}, \bibinfo
  {author} {\bibfnamefont {A.}~\bibnamefont {Ourjoumtsev}}, \ and\ \bibinfo
  {author} {\bibfnamefont {P.}~\bibnamefont {Grangier}},\ }\href {\doibase
  10.1103/PhysRevA.88.053845} {\bibfield  {journal} {\bibinfo  {journal} {Phys.
  Rev. A}\ }\textbf {\bibinfo {volume} {88}},\ \bibinfo {pages} {053845}
  (\bibinfo {year} {2013})}\BibitemShut {NoStop}%
\bibitem [{\citenamefont {Grankin}\ \emph
  {et~al.}(2014{\natexlab{a}})\citenamefont {Grankin}, \citenamefont {Brion},
  \citenamefont {Bimbard}, \citenamefont {Boddeda}, \citenamefont {Usmani},
  \citenamefont {Ourjoumtsev},\ and\ \citenamefont {Grangier}}]{Grangier2014}%
  \BibitemOpen
  \bibfield  {author} {\bibinfo {author} {\bibfnamefont {A.}~\bibnamefont
  {Grankin}}, \bibinfo {author} {\bibfnamefont {E.}~\bibnamefont {Brion}},
  \bibinfo {author} {\bibfnamefont {E.}~\bibnamefont {Bimbard}}, \bibinfo
  {author} {\bibfnamefont {R.}~\bibnamefont {Boddeda}}, \bibinfo {author}
  {\bibfnamefont {I.}~\bibnamefont {Usmani}}, \bibinfo {author} {\bibfnamefont
  {A.}~\bibnamefont {Ourjoumtsev}}, \ and\ \bibinfo {author} {\bibfnamefont
  {P.}~\bibnamefont {Grangier}},\ }\href
  {http://stacks.iop.org/1367-2630/16/i=4/a=043020} {\bibfield  {journal}
  {\bibinfo  {journal} {New Journal of Physics}\ }\textbf {\bibinfo {volume}
  {16}},\ \bibinfo {pages} {043020} (\bibinfo {year}
  {2014}{\natexlab{a}})}\BibitemShut {NoStop}%
\bibitem [{\citenamefont {Grankin}\ \emph
  {et~al.}(2014{\natexlab{b}})\citenamefont {Grankin}, \citenamefont {Brion},
  \citenamefont {Bimbard}, \citenamefont {Boddeda}, \citenamefont {Usmani},
  \citenamefont {Ourjoumtsev},\ and\ \citenamefont {Grangier}}]{grankin2014}%
  \BibitemOpen
  \bibfield  {author} {\bibinfo {author} {\bibfnamefont {A.}~\bibnamefont
  {Grankin}}, \bibinfo {author} {\bibfnamefont {E.}~\bibnamefont {Brion}},
  \bibinfo {author} {\bibfnamefont {E.}~\bibnamefont {Bimbard}}, \bibinfo
  {author} {\bibfnamefont {R.}~\bibnamefont {Boddeda}}, \bibinfo {author}
  {\bibfnamefont {I.}~\bibnamefont {Usmani}}, \bibinfo {author} {\bibfnamefont
  {A.}~\bibnamefont {Ourjoumtsev}}, \ and\ \bibinfo {author} {\bibfnamefont
  {P.}~\bibnamefont {Grangier}},\ }\href {\doibase
  10.1088/1367-2630/16/4/043020} {\bibfield  {journal} {\bibinfo  {journal}
  {New Journal of Physics}\ }\textbf {\bibinfo {volume} {16}},\ \bibinfo
  {pages} {1} (\bibinfo {year} {2014}{\natexlab{b}})},\ \Eprint
  {http://arxiv.org/abs/arXiv:1312.2621v1} {arXiv:arXiv:1312.2621v1}
  \BibitemShut {NoStop}%
\bibitem [{\citenamefont {Grankin}\ \emph {et~al.}(2015)\citenamefont
  {Grankin}, \citenamefont {Brion}, \citenamefont {Bimbard}, \citenamefont
  {Boddeda}, \citenamefont {Usmani}, \citenamefont {Ourjoumtsev},\ and\
  \citenamefont {Grangier}}]{Grankin2015}%
  \BibitemOpen
  \bibfield  {author} {\bibinfo {author} {\bibfnamefont {A.}~\bibnamefont
  {Grankin}}, \bibinfo {author} {\bibfnamefont {E.}~\bibnamefont {Brion}},
  \bibinfo {author} {\bibfnamefont {E.}~\bibnamefont {Bimbard}}, \bibinfo
  {author} {\bibfnamefont {R.}~\bibnamefont {Boddeda}}, \bibinfo {author}
  {\bibfnamefont {I.}~\bibnamefont {Usmani}}, \bibinfo {author} {\bibfnamefont
  {A.}~\bibnamefont {Ourjoumtsev}}, \ and\ \bibinfo {author} {\bibfnamefont
  {P.}~\bibnamefont {Grangier}},\ }\href {\doibase 10.1103/PhysRevA.92.043841}
  {\bibfield  {journal} {\bibinfo  {journal} {Phys. Rev. A}\ }\textbf {\bibinfo
  {volume} {92}},\ \bibinfo {pages} {043841} (\bibinfo {year}
  {2015})}\BibitemShut {NoStop}%
\bibitem [{\citenamefont {Bienias}\ and\ \citenamefont
  {Büchler}(2016)}]{Bienias2016}%
  \BibitemOpen
  \bibfield  {author} {\bibinfo {author} {\bibfnamefont {P.}~\bibnamefont
  {Bienias}}\ and\ \bibinfo {author} {\bibfnamefont {H.~P.}\ \bibnamefont
  {Büchler}},\ }\href {http://stacks.iop.org/1367-2630/18/i=12/a=123026}
  {\bibfield  {journal} {\bibinfo  {journal} {New Journal of Physics}\ }\textbf
  {\bibinfo {volume} {18}},\ \bibinfo {pages} {123026} (\bibinfo {year}
  {2016})}\BibitemShut {NoStop}%
\bibitem [{\citenamefont {Pritchard}\ \emph {et~al.}(2010)\citenamefont
  {Pritchard}, \citenamefont {Maxwell}, \citenamefont {Gauguet}, \citenamefont
  {Weatherill}, \citenamefont {Jones},\ and\ \citenamefont
  {Adams}}]{Pritchard2010}%
  \BibitemOpen
  \bibfield  {author} {\bibinfo {author} {\bibfnamefont {J.~D.}\ \bibnamefont
  {Pritchard}}, \bibinfo {author} {\bibfnamefont {D.}~\bibnamefont {Maxwell}},
  \bibinfo {author} {\bibfnamefont {A.}~\bibnamefont {Gauguet}}, \bibinfo
  {author} {\bibfnamefont {K.~J.}\ \bibnamefont {Weatherill}}, \bibinfo
  {author} {\bibfnamefont {M.~P.~A.}\ \bibnamefont {Jones}}, \ and\ \bibinfo
  {author} {\bibfnamefont {C.~S.}\ \bibnamefont {Adams}},\ }\href {\doibase
  10.1103/PhysRevLett.105.193603} {\bibfield  {journal} {\bibinfo  {journal}
  {Phys. Rev. Lett.}\ }\textbf {\bibinfo {volume} {105}},\ \bibinfo {pages}
  {193603} (\bibinfo {year} {2010})}\BibitemShut {NoStop}%
\bibitem [{\citenamefont {Pritchard}\ \emph {et~al.}(2011)\citenamefont
  {Pritchard}, \citenamefont {Gauguet}, \citenamefont {Weatherill},\ and\
  \citenamefont {Adams}}]{Pritchard2011a}%
  \BibitemOpen
  \bibfield  {author} {\bibinfo {author} {\bibfnamefont {J.~D.}\ \bibnamefont
  {Pritchard}}, \bibinfo {author} {\bibfnamefont {A.}~\bibnamefont {Gauguet}},
  \bibinfo {author} {\bibfnamefont {K.~J.}\ \bibnamefont {Weatherill}}, \ and\
  \bibinfo {author} {\bibfnamefont {C.~S.}\ \bibnamefont {Adams}},\ }\href
  {http://stacks.iop.org/0953-4075/44/i=18/a=184019} {\bibfield  {journal}
  {\bibinfo  {journal} {Journal of Physics B: Atomic, Molecular and Optical
  Physics}\ }\textbf {\bibinfo {volume} {44}},\ \bibinfo {pages} {184019}
  (\bibinfo {year} {2011})}\BibitemShut {NoStop}%
\bibitem [{\citenamefont {Boddeda}\ \emph {et~al.}(2016)\citenamefont
  {Boddeda}, \citenamefont {Usmani}, \citenamefont {Bimbard}, \citenamefont
  {Grankin}, \citenamefont {Ourjoumtsev}, \citenamefont {Brion},\ and\
  \citenamefont {Grangier}}]{Boddeda2016Experiment}%
  \BibitemOpen
  \bibfield  {author} {\bibinfo {author} {\bibfnamefont {R.}~\bibnamefont
  {Boddeda}}, \bibinfo {author} {\bibfnamefont {I.}~\bibnamefont {Usmani}},
  \bibinfo {author} {\bibfnamefont {E.}~\bibnamefont {Bimbard}}, \bibinfo
  {author} {\bibfnamefont {A.}~\bibnamefont {Grankin}}, \bibinfo {author}
  {\bibfnamefont {A.}~\bibnamefont {Ourjoumtsev}}, \bibinfo {author}
  {\bibfnamefont {E.}~\bibnamefont {Brion}}, \ and\ \bibinfo {author}
  {\bibfnamefont {P.}~\bibnamefont {Grangier}},\ }\href
  {http://stacks.iop.org/0953-4075/49/i=8/a=084005} {\bibfield  {journal}
  {\bibinfo  {journal} {Journal of Physics B: Atomic, Molecular and Optical
  Physics}\ }\textbf {\bibinfo {volume} {49}},\ \bibinfo {pages} {084005}
  (\bibinfo {year} {2016})}\BibitemShut {NoStop}%
\bibitem [{\citenamefont {Parigi}\ \emph {et~al.}(2012)\citenamefont {Parigi},
  \citenamefont {Bimbard}, \citenamefont {Stanojevic}, \citenamefont
  {Hilliard}, \citenamefont {Nogrette}, \citenamefont {Tualle-Brouri},
  \citenamefont {Ourjoumtsev},\ and\ \citenamefont {Grangier}}]{Parigi2012}%
  \BibitemOpen
  \bibfield  {author} {\bibinfo {author} {\bibfnamefont {V.}~\bibnamefont
  {Parigi}}, \bibinfo {author} {\bibfnamefont {E.}~\bibnamefont {Bimbard}},
  \bibinfo {author} {\bibfnamefont {J.}~\bibnamefont {Stanojevic}}, \bibinfo
  {author} {\bibfnamefont {A.~J.}\ \bibnamefont {Hilliard}}, \bibinfo {author}
  {\bibfnamefont {F.}~\bibnamefont {Nogrette}}, \bibinfo {author}
  {\bibfnamefont {R.}~\bibnamefont {Tualle-Brouri}}, \bibinfo {author}
  {\bibfnamefont {A.}~\bibnamefont {Ourjoumtsev}}, \ and\ \bibinfo {author}
  {\bibfnamefont {P.}~\bibnamefont {Grangier}},\ }\href {\doibase
  10.1103/PhysRevLett.109.233602} {\bibfield  {journal} {\bibinfo  {journal}
  {Phys. Rev. Lett.}\ }\textbf {\bibinfo {volume} {109}},\ \bibinfo {pages}
  {233602} (\bibinfo {year} {2012})}\BibitemShut {NoStop}%
\bibitem [{\citenamefont {Singer}\ \emph {et~al.}(2005)\citenamefont {Singer},
  \citenamefont {Stanojevic}, \citenamefont {Weidemüller},\ and\ \citenamefont
  {Côté}}]{Singer2005}%
  \BibitemOpen
  \bibfield  {author} {\bibinfo {author} {\bibfnamefont {K.}~\bibnamefont
  {Singer}}, \bibinfo {author} {\bibfnamefont {J.}~\bibnamefont {Stanojevic}},
  \bibinfo {author} {\bibfnamefont {M.}~\bibnamefont {Weidemüller}}, \ and\
  \bibinfo {author} {\bibfnamefont {R.}~\bibnamefont {Côté}},\ }\href
  {http://stacks.iop.org/0953-4075/38/i=2/a=021} {\bibfield  {journal}
  {\bibinfo  {journal} {Journal of Physics B: Atomic, Molecular and Optical
  Physics}\ }\textbf {\bibinfo {volume} {38}},\ \bibinfo {pages} {S295}
  (\bibinfo {year} {2005})}\BibitemShut {NoStop}%
\bibitem [{\citenamefont {Li}\ \emph {et~al.}(2003)\citenamefont {Li},
  \citenamefont {Mourachko}, \citenamefont {Noel},\ and\ \citenamefont
  {Gallagher}}]{Wenhui2003}%
  \BibitemOpen
  \bibfield  {author} {\bibinfo {author} {\bibfnamefont {W.}~\bibnamefont
  {Li}}, \bibinfo {author} {\bibfnamefont {I.}~\bibnamefont {Mourachko}},
  \bibinfo {author} {\bibfnamefont {M.~W.}\ \bibnamefont {Noel}}, \ and\
  \bibinfo {author} {\bibfnamefont {T.~F.}\ \bibnamefont {Gallagher}},\ }\href
  {\doibase 10.1103/PhysRevA.67.052502} {\bibfield  {journal} {\bibinfo
  {journal} {Phys. Rev. A}\ }\textbf {\bibinfo {volume} {67}},\ \bibinfo
  {pages} {052502} (\bibinfo {year} {2003})}\BibitemShut {NoStop}%
\bibitem [{\citenamefont {Parigi}\ \emph {et~al.}()\citenamefont {Parigi},
  \citenamefont {Bimbard}, \citenamefont {Stanojevic}, \citenamefont
  {Hilliard}, \citenamefont {Nogrette}, \citenamefont {Tualle-Brouri},
  \citenamefont {Ourjoumtsev},\ and\ \citenamefont {Grangier}}]{Parigi}%
  \BibitemOpen
  \bibfield  {author} {\bibinfo {author} {\bibfnamefont {V.}~\bibnamefont
  {Parigi}}, \bibinfo {author} {\bibfnamefont {E.}~\bibnamefont {Bimbard}},
  \bibinfo {author} {\bibfnamefont {J.}~\bibnamefont {Stanojevic}}, \bibinfo
  {author} {\bibfnamefont {A.~J.}\ \bibnamefont {Hilliard}}, \bibinfo {author}
  {\bibfnamefont {F.}~\bibnamefont {Nogrette}}, \bibinfo {author}
  {\bibfnamefont {R.}~\bibnamefont {Tualle-Brouri}}, \bibinfo {author}
  {\bibfnamefont {A.}~\bibnamefont {Ourjoumtsev}}, \ and\ \bibinfo {author}
  {\bibfnamefont {P.}~\bibnamefont {Grangier}},\ }\href {\doibase
  10.1103/PhysRevLett.109.233602} {\
  10.1103/PhysRevLett.109.233602}\BibitemShut {NoStop}%
\bibitem [{\citenamefont {Ga{\"{e}}tan}\ \emph {et~al.}(2009)\citenamefont
  {Ga{\"{e}}tan}, \citenamefont {Miroshnychenko}, \citenamefont {Wilk},
  \citenamefont {Chotia}, \citenamefont {Viteau}, \citenamefont {Comparat},
  \citenamefont {Pillet}, \citenamefont {Browaeys},\ and\ \citenamefont
  {Grangier}}]{Gaetan2009}%
  \BibitemOpen
  \bibfield  {author} {\bibinfo {author} {\bibfnamefont {A.}~\bibnamefont
  {Ga{\"{e}}tan}}, \bibinfo {author} {\bibfnamefont {Y.}~\bibnamefont
  {Miroshnychenko}}, \bibinfo {author} {\bibfnamefont {T.}~\bibnamefont
  {Wilk}}, \bibinfo {author} {\bibfnamefont {A.}~\bibnamefont {Chotia}},
  \bibinfo {author} {\bibfnamefont {M.}~\bibnamefont {Viteau}}, \bibinfo
  {author} {\bibfnamefont {D.}~\bibnamefont {Comparat}}, \bibinfo {author}
  {\bibfnamefont {P.}~\bibnamefont {Pillet}}, \bibinfo {author} {\bibfnamefont
  {A.}~\bibnamefont {Browaeys}}, \ and\ \bibinfo {author} {\bibfnamefont
  {P.}~\bibnamefont {Grangier}},\ }\href
  {https://www.nature.com/articles/nphys1183} {\bibfield  {journal} {\bibinfo
  {journal} {Nature Physics}\ }\textbf {\bibinfo {volume} {5}} (\bibinfo {year}
  {2009})}\BibitemShut {NoStop}%
\bibitem [{\citenamefont {Abel}\ \emph {et~al.}(2009)\citenamefont {Abel},
  \citenamefont {Mohapatra}, \citenamefont {Bason}, \citenamefont {Pritchard},
  \citenamefont {Weatherill}, \citenamefont {Raitzsch},\ and\ \citenamefont
  {Adams}}]{Abel2009}%
  \BibitemOpen
  \bibfield  {author} {\bibinfo {author} {\bibfnamefont {R.~P.}\ \bibnamefont
  {Abel}}, \bibinfo {author} {\bibfnamefont {A.~K.}\ \bibnamefont {Mohapatra}},
  \bibinfo {author} {\bibfnamefont {M.~G.}\ \bibnamefont {Bason}}, \bibinfo
  {author} {\bibfnamefont {J.~D.}\ \bibnamefont {Pritchard}}, \bibinfo {author}
  {\bibfnamefont {K.~J.}\ \bibnamefont {Weatherill}}, \bibinfo {author}
  {\bibfnamefont {U.}~\bibnamefont {Raitzsch}}, \ and\ \bibinfo {author}
  {\bibfnamefont {C.~S.}\ \bibnamefont {Adams}},\ }\href {\doibase
  10.1063/1.3086305} {\bibfield  {journal} {\bibinfo  {journal} {Applied
  Physics Letters}\ }\textbf {\bibinfo {volume} {94}},\ \bibinfo {pages}
  {071107} (\bibinfo {year} {2009})}\BibitemShut {NoStop}%
\bibitem [{Note1()}]{Note1}%
  \BibitemOpen
  \bibinfo {note} {For simplicity we have assumed that the probability of
  Rydberg excitations is low enough to neglect any spatial correlations in the
  Rydberg density}\BibitemShut {NoStop}%
\bibitem [{\citenamefont {Pritchard}(2011)}]{Pritchard2011b}%
  \BibitemOpen
  \bibfield  {author} {\bibinfo {author} {\bibfnamefont {J.~D.}\ \bibnamefont
  {Pritchard}},\ }\bibfield  {booktitle} {\emph {\bibinfo {booktitle}
  {{Cooperative Optical Non-linearity in a blockaded Rydberg Ensemble}}},\
  }\href {http://etheses.dur.ac.uk/782/} {\  (\bibinfo {year}
  {2011})}\BibitemShut {NoStop}%
\bibitem [{\citenamefont {Boulier}\ \emph {et~al.}(2017)\citenamefont
  {Boulier}, \citenamefont {Magnan}, \citenamefont {Bracamontes}, \citenamefont
  {Maslek}, \citenamefont {Goldschmidt}, \citenamefont {Young}, \citenamefont
  {Gorshkov}, \citenamefont {Rolston},\ and\ \citenamefont
  {Porto}}]{Goldschmidt2017}%
  \BibitemOpen
  \bibfield  {author} {\bibinfo {author} {\bibfnamefont {T.}~\bibnamefont
  {Boulier}}, \bibinfo {author} {\bibfnamefont {E.}~\bibnamefont {Magnan}},
  \bibinfo {author} {\bibfnamefont {C.}~\bibnamefont {Bracamontes}}, \bibinfo
  {author} {\bibfnamefont {J.}~\bibnamefont {Maslek}}, \bibinfo {author}
  {\bibfnamefont {E.~A.}\ \bibnamefont {Goldschmidt}}, \bibinfo {author}
  {\bibfnamefont {J.~T.}\ \bibnamefont {Young}}, \bibinfo {author}
  {\bibfnamefont {A.~V.}\ \bibnamefont {Gorshkov}}, \bibinfo {author}
  {\bibfnamefont {S.~L.}\ \bibnamefont {Rolston}}, \ and\ \bibinfo {author}
  {\bibfnamefont {J.~V.}\ \bibnamefont {Porto}},\ }\href {\doibase
  10.1103/PhysRevA.96.053409} {\bibfield  {journal} {\bibinfo  {journal} {Phys.
  Rev. A}\ }\textbf {\bibinfo {volume} {96}},\ \bibinfo {pages} {053409}
  (\bibinfo {year} {2017})}\BibitemShut {NoStop}%
\bibitem [{\citenamefont {Dmochowski}(2016)}]{Dmochowski2012}%
  \BibitemOpen
  \bibfield  {author} {\bibinfo {author} {\bibfnamefont {G.}~\bibnamefont
  {Dmochowski}},\ }\bibfield  {booktitle} {\emph {\bibinfo {booktitle} {{8-Port
  Homodyne Detection of EIT-Enhanced Cross-Phase Modulation Using Broadband
  Signal Pulses}}},\ }\href
  {https://tspace.library.utoronto.ca/bitstream/1807/72963/3/Dmochowski_Greg_201606_PhD_thesis.pdf}
  {\  (\bibinfo {year} {2016})}\BibitemShut {NoStop}%
\bibitem [{\citenamefont {Feizpour}\ \emph {et~al.}(2016)\citenamefont
  {Feizpour}, \citenamefont {Dmochowski},\ and\ \citenamefont
  {Steinberg}}]{Feizpour2016d}%
  \BibitemOpen
  \bibfield  {author} {\bibinfo {author} {\bibfnamefont {A.}~\bibnamefont
  {Feizpour}}, \bibinfo {author} {\bibfnamefont {G.}~\bibnamefont
  {Dmochowski}}, \ and\ \bibinfo {author} {\bibfnamefont {A.~M.}\ \bibnamefont
  {Steinberg}},\ }\href {\doibase 10.1103/PhysRevA.93.013834} {\bibfield
  {journal} {\bibinfo  {journal} {Phys. Rev. A}\ }\textbf {\bibinfo {volume}
  {93}},\ \bibinfo {pages} {013834} (\bibinfo {year} {2016})}\BibitemShut
  {NoStop}%
\bibitem [{\citenamefont {Boyd}(2008)}]{Boyd3rd}%
  \BibitemOpen
  \bibfield  {author} {\bibinfo {author} {\bibfnamefont {R.~W.}\ \bibnamefont
  {Boyd}},\ }\href@noop {} {\emph {\bibinfo {title} {Nonlinear Optics, Third
  Edition}}},\ \bibinfo {edition} {3rd}\ ed.\ (\bibinfo  {publisher} {Academic
  Press, Inc.},\ \bibinfo {address} {Orlando, FL, USA},\ \bibinfo {year}
  {2008})\BibitemShut {NoStop}%
\bibitem [{\citenamefont {Venkataraman}\ \emph {et~al.}(2013)\citenamefont
  {Venkataraman}, \citenamefont {Saha},\ and\ \citenamefont
  {Gaeta}}]{Venkataraman2013}%
  \BibitemOpen
  \bibfield  {author} {\bibinfo {author} {\bibfnamefont {V.}~\bibnamefont
  {Venkataraman}}, \bibinfo {author} {\bibfnamefont {K.}~\bibnamefont {Saha}},
  \ and\ \bibinfo {author} {\bibfnamefont {A.~L.}\ \bibnamefont {Gaeta}},\
  }\href {\doibase 10.1038/nphoton.2012.283} {\bibfield  {journal} {\bibinfo
  {journal} {Nature Photonics}\ }\textbf {\bibinfo {volume} {7}},\ \bibinfo
  {pages} {138} (\bibinfo {year} {2013})}\BibitemShut {NoStop}%
\bibitem [{\citenamefont {Hau}\ \emph {et~al.}(1999)\citenamefont {Hau},
  \citenamefont {Harris}, \citenamefont {Dutton},\ and\ \citenamefont
  {Behroozi}}]{Hau1999a}%
  \BibitemOpen
  \bibfield  {author} {\bibinfo {author} {\bibfnamefont {L.~V.}\ \bibnamefont
  {Hau}}, \bibinfo {author} {\bibfnamefont {S.~E.}\ \bibnamefont {Harris}},
  \bibinfo {author} {\bibfnamefont {Z.}~\bibnamefont {Dutton}}, \ and\ \bibinfo
  {author} {\bibfnamefont {C.~H.}\ \bibnamefont {Behroozi}},\ }\href
  {https://www.nature.com/articles/17561.pdf} {\bibfield  {journal} {\bibinfo
  {journal} {Nature}\ }\textbf {\bibinfo {volume} {397}},\ \bibinfo {pages}
  {594} (\bibinfo {year} {1999})}\BibitemShut {NoStop}%
\bibitem [{\citenamefont {Feizpour}\ \emph {et~al.}(2015)\citenamefont
  {Feizpour}, \citenamefont {Hallaji}, \citenamefont {Dmochowski},\ and\
  \citenamefont {Steinberg}}]{Feizpour2015}%
  \BibitemOpen
  \bibfield  {author} {\bibinfo {author} {\bibfnamefont {A.}~\bibnamefont
  {Feizpour}}, \bibinfo {author} {\bibfnamefont {M.}~\bibnamefont {Hallaji}},
  \bibinfo {author} {\bibfnamefont {G.}~\bibnamefont {Dmochowski}}, \ and\
  \bibinfo {author} {\bibfnamefont {A.~M.}\ \bibnamefont {Steinberg}},\ }\href
  {https://doi.org/10.1038/nphys3433} {\bibfield  {journal} {\bibinfo
  {journal} {Nature Physics}\ }\textbf {\bibinfo {volume} {11}},\ \bibinfo
  {pages} {905} (\bibinfo {year} {2015})}\BibitemShut {NoStop}%
\end{thebibliography}%

\end{document}